\documentclass{amsart}
\usepackage{amsmath,graphicx}

\usepackage{amsthm}
\usepackage{amssymb}
\usepackage{xypic}
\usepackage[square]{natbib}
\setcitestyle{numbers}
\theoremstyle{plain}

\theoremstyle{definition}

\title[Finding non-compliant documents using error messages]{Looking for non-compliant documents using error messages from multiple parsers}
\author{Michael Robinson}
\address{Mathematics and Statistics\\
American University\\
Washington, DC, USA}
\email{michaelr@american.edu}

\begin{document}

\begin{abstract}
Whether a file is accepted by a single parser is not a reliable indication of whether a file complies with its stated format.  Bugs within both the parser and the format specification mean that a compliant file may fail to parse, or that a non-compliant file might be read without any apparent trouble.  The latter situation presents a significant security risk, and should be avoided.  This article suggests that a better way to assess format specification compliance is to examine the set of error messages produced by a set of parsers rather than a single parser.  If both a sample of compliant files and a sample of non-compliant files are available, then we show how a statistical test based on a pseudo-likelihood ratio can be very effective at determining a file's compliance.  Our method is format agnostic, and does not directly rely upon a formal specification of the format.  Although this article focuses upon the case of the PDF format (ISO 32000-2), we make no attempt to use any specific details of the format.  Furthermore, we show how principal components analysis can be useful for a format specification designer to assess the quality and structure of these samples of files and parsers.   While these tests are absolutely rudimentary, it appears that their use to measure file format variability and to identify non-compliant files is both novel and surprisingly effective. 
\end{abstract}

\maketitle


\section{Introduction}

Modern file formats are often quite complex, yet the formal specifications for some common formats can be ambiguous or confusing.  A single parser is therefore not a reliable arbiter of file format compliance: it may incorrectly deem a compliant file as non-compliant, or conversely it may parse a non-compliant file (perhaps with disastrous consequences).  For widely used file formats, there are usually several readily available parsers.  It is natural to ask if a statistical approach that leverages multiple existing parsers -- but is otherwise format agnostic -- might suffice to discriminate between compliant and non-compliant files.  

This article describes an exploratory technique and a statistical test for identifying files whose parser behavior is unusual.  The techniques presented perform no direct inspection of the contents of any file.  Certainly the content of a given file plays an important role in its usage, but the techniques of this article only ``see the content'' through the lens of an ensemble of parsers.  Our techniques are therefore also well-suited to assessing the background variability of parser behavior on different classes of files.  Since our approach aims to leverage existing parsers in their unmodified, uninstrumented state, the statistical techniques could be used on many different file formats without much alteration.

For the purpose of this article, a \emph{file format} consists of a set of compliant files and a set of non-compliant files.   Formal specifications specify properties that compliant files must have, but formal specifications need not be present for there to be an agreed-upon file format.
This article presents a new statistical test, which we call \emph{the Bernoulli misclassification test}, that determines whether a given file is more representative of the compliant files or of the non-compliant files.
In order to perform such a test, we require \emph{samples} of both sets: namely a sample of compliant and a sample of non-compliant files.
Because realistic samples of files are large and difficult to curate, the sample of compliant files may be contaminated with files that should not be considered as compliant.
Conversely, there may be some files in our sample that are erroneously marked as non-compliant.
Our statistical test is designed to identify these \emph{misclassified} files.

The foundation of any statistical approach necessarily relies on both data coverage and sufficient sampling to ensure good estimation of the relevant governing parameters.  Our approach here is no different, as the basis of the statistical test relies upon parameters estimated from the data in order to be effective.  Given that our approach is format agnostic, it is reliant upon not only a good sample of files but also a good sample of parsers. 

To test our approach, this article presents a case study using the PDF specification (ISO 32000-2), because there are many extant open source parsers with distinct underlying codebases.
The test data presented in this article were developed by an independent test and evaluation team in support of the ``DARPA SafeDocs Evaluation 2'' exercise.  The data consist of two datasets corresponding to the samples explained above: a sample of largely compliant files and a sample of largely non-compliant files.  Each of the files (in both samples) was manually tested for compliance, so that the performance of our statistical test could be determined.  While the fraction of misclassified files in the two datasets differ, the two datasets were sufficiently clean so as to allow reliable enough parameter estimation for our statistical test.

While the statistical methods discussed in this article are absolutely rudimentary, they did locate files that were truly misclassified with surprising effectiveness.  Although these statistical methods surely do not suffice on their own for all purposes, they are easy to deploy and understand.  We suggest that they ought to be part of the format hacker's toolbox.

\section{Historical context}

There appears to be very little work in analyzing file format compliance using statistical tools.
In contrast to what we present here, most format compliance assessment that the author is aware of is performed using techniques common in compiler theory.  For instance, \cite{aho1986compilers,schipka2009detection} explain the typical approaches.

The closest connections to this article appear to be various techniques for identifying malware using the structure of file \emph{contents} rather than responses to those contents.  For instance \cite{belaoued2015real} looked for statistical features characteristic of malware present in headers of executable files.  Using the structure of file contents, ransomware can be classified statistically \cite{al2018ransomware, 8685181}.  Similar to our use of error messages on files, the distribution of API calls can be useful in identifying malware as it executes \cite{ALAZAB201591}.  Other behavioral indicators, such as performance counters \cite{demme2013feasibility} can be useful as well.  However, it appears that the use of statistical tools to determine file format compliance is completely unanticipated and novel.

In a few cases, statistical methods are useful in identifying file formats that might be difficult to archive or curate \cite{graf2013risk,Pearson_2008,lawrence2000risk}.  

\section{Data description}

This article focuses on the analysis of PDF files, whose format is determined by the PDF specification (ISO 32000-2).
It is recognized that the specification is ambiguous in places, and that there are many proprietary extensions to the specification.
Because of this, many ``PDF files'' may not be completely compliant or may be close enough to compliance to parse correctly.
On the other hand, bugs within the parsers may cause them to accept non-compliant files.
To explore these issues, the test and evaluation team collected a corpus of ``PDF files''
into two datasets: {\tt internet\_sourced} and {\tt dangerous} comprising a total of $10000$ files.

Within each dataset, the files were manually determined to be either ``valid'', that is that they are compliant with the PDF specification, or ``rejected'', which means that they fail to comply with the specification.
In this article, we treat the ``valid'' or ``rejected'' determinations as experimental \emph{ground truth} for the files.
Since our Bernoulli misclassification test did not use these determinations, we were able to use them to estimate the test's accuracy (discussed in Section \ref{sec:bern}).
The overall statistics of both datasets are shown in Table \ref{tab:datasets}.
A standard $\chi^2$ test reveals that the differences in compliance between the two datasets is statistically significant ($p < 0.0001$).
Although it is far from true, we took the {\tt internet\_sourced} set to be our sample of predominantly compliant files,
and we took the {\tt dangerous} set to be our sample of predominantly non-compliant files.
The significant difference between the two samples is precisely what our misclassification test leverages in order to find misclassified files.

\begin{table}
  \begin{center}
    \caption{Counts of files in the {\tt internet\_sourced} and {\tt dangerous} datasets}
    \label{tab:datasets}
    \begin{tabular}{|l|c|c||c|}
      \hline
      Dataset & Valid & Rejected & Total\\
      \hline
      \hline
          {\tt internet\_sourced}&7206&1794&9000\\
          \hline
              {\tt dangerous}&488&516&1000\\
              \hline
              \hline
              Totals&7694&2306&10000\\
              \hline
    \end{tabular}
  \end{center}
\end{table}

Rather than looking at the contents of each file, we reasoned that there are already many extant parsers that do just that.
Therefore, we ran each file through each of a large collection of parsers shown in Table \ref{tab:messages}.
We selected these parsers based on their easy availability and with the understanding that many of them do not share code.
This latter fact ensures that places within the PDF specification that are ambiguous may receive several interpretations by different parsers.
The output to {\tt stderr} from each parser was collected for each file, and a set of $955$ regular expressions were used to identify which error messages had occurred for each parser-file pair (see Table \ref{tab:messages}).

As an example, the $1000$-th file in the {\tt internet\_sourced} dataset was considered ``valid,'' yet produced $7$ distinct messages:
\begin{description}
\item[58] {\tt caradoc\_extract}: {\tt Type error : Invalid variant type},
\item[254] {\tt caradoc\_stats}: {\tt Type error : Invalid variant type},
\item[393] {\tt caradoc\_stats\_strict}: {\tt PDF error : Syntax error},
\item[589] {\tt hammer}: uncategorized error,
\item[683] {\tt pdfium}: uncategorized error,
\item[910] {\tt xpdf\_pdfinfo}: uncategorized error, and
\item[911] {\tt xpdf\_pdftops}: uncategorized error.
\end{description}
It is important to recognize that our method \emph{makes no attempt to interpret} the semantic meanings of these error messages.  Instead, we are merely interested in the statistics of the co-occurrence of these messages.

The data can be tabulated as an integer \emph{relation matrix}, in which each row corresponds to a particular regular expression (an \emph{error message} in what follows) and each column corresponds to a particular file.  Each entry records the number of times a given error message occurred for each file.  Return values to the operating system were not collected, though if desired these could have simply been added as additional ``messages'' as rows in the relation matrix.

\begin{table}
  \begin{center}
    \caption{Error message counts and rows per parser}
    \label{tab:messages}
    \begin{tabular}{|l|c|c|c|}
      \hline
      Parser&First row&Last row&Total message regexes\\
      \hline
      {\tt caradoc\_extract}&1&196&196\\
{\tt caradoc\_stats}&197&392&196\\
{\tt caradoc\_stats\_strict}&393&588&196\\
{\tt hammer}&589&589&1\\
{\tt mutool\_show}&590&635&46\\
{\tt mutool\_clean}&636&681&46\\
{\tt origami\_pdfcop}&682&682&1\\
{\tt pdfium}&683&683&1\\
{\tt pdfminer\_dumppdf}&684&703&20\\
{\tt pdfminer\_pdf2txt}&704&723&20\\
{\tt pdftk\_server}&724&724&1\\
{\tt pdftools\_pdfid}&725&729&5\\
{\tt pdftools\_pdfparser}&730&734&5\\
{\tt peepdf}&735&735&1\\
{\tt poppler\_pdfinfo}&736&792&57\\
{\tt poppler\_pdftocairo}&793&849&57\\
{\tt poppler\_pdftops}&850&906&57\\
{\tt qpdf}&907&907&1\\
{\tt verapdf\_greenfield}&908&908&1\\
{\tt verapdf\_pdfbox}&909&909&1\\
{\tt xpdf\_pdfinfo}&910&910&1\\
{\tt xpdf\_pdftops}&911&911&1\\
{\tt caradoc\_extract}&912&913&2\\
{\tt caradoc\_stats}&914&915&2\\
{\tt caradoc\_stats\_strict}&916&917&2\\
{\tt hammer}&918&919&2\\
{\tt mutool\_clean}&920&921&2\\
{\tt mutool\_show}&922&923&2\\
{\tt origami\_pdfcop}&924&925&2\\
{\tt pdfium}&926&927&2\\
{\tt pdfminer\_dumppdf}&928&929&2\\
{\tt pdfminer\_pdf2txt}&930&931&2\\
{\tt pdftk\_server}&932&933&2\\
{\tt pdftools\_pdfid}&934&935&2\\
{\tt pdftools\_pdfparser}&936&937&2\\
{\tt peepdf}&938&939&2\\
{\tt poppler\_pdfinfo}&940&941&2\\
{\tt poppler\_pdftocairo}&942&943&2\\
{\tt poppler\_pdftops}&944&945&2\\
{\tt qpdf}&946&947&2\\
{\tt verapdf\_greenfield}&948&949&2\\
{\tt verapdf\_pdfbox}&950&951&2\\
{\tt xpdf\_pdfinfo}&952&953&2\\
      \hline
      Total&&&955\\
      \hline
    \end{tabular}
  \end{center}
\end{table}

\begin{figure}
  \begin{center}
    \includegraphics[width=5in]{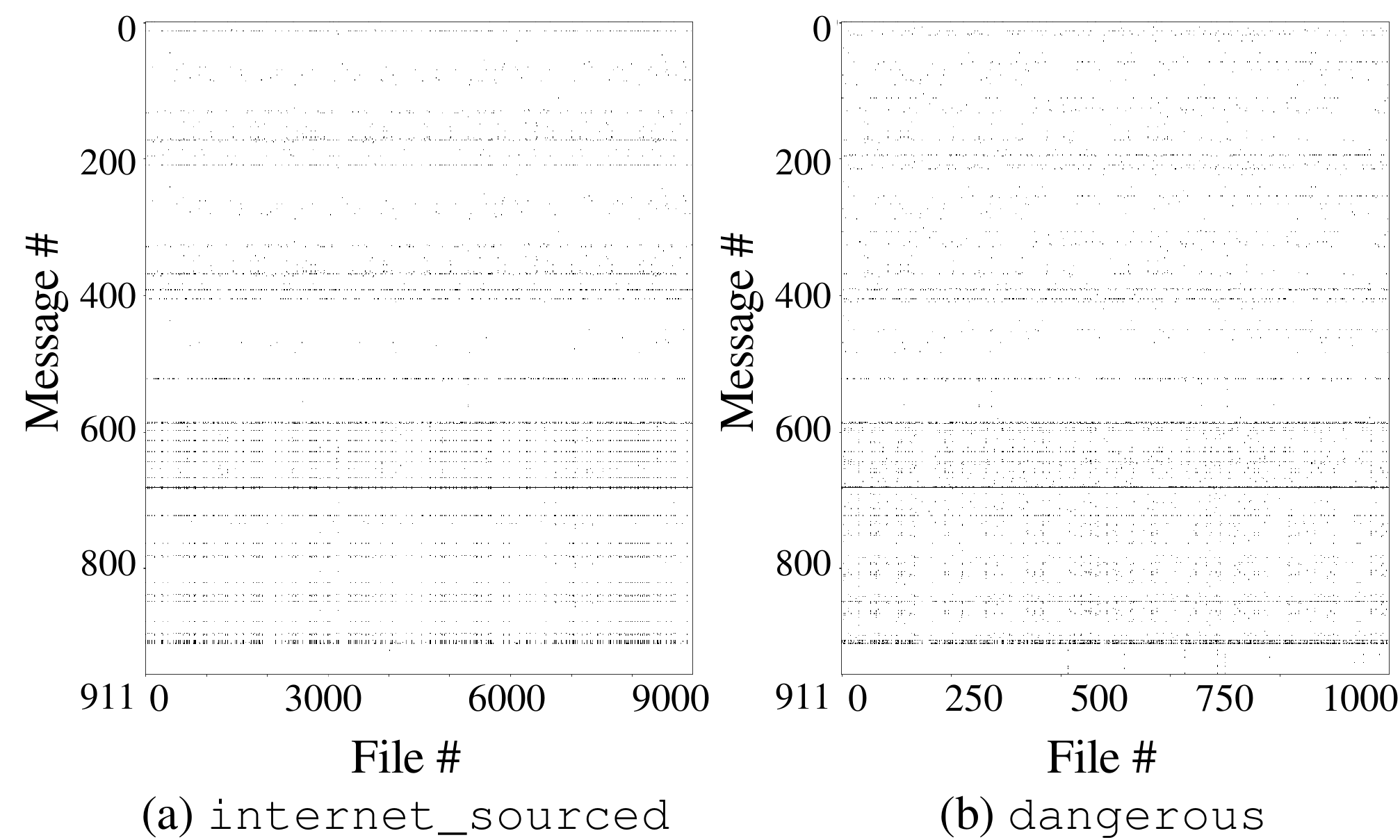}
    \caption{The relation matrix for (a) {\tt internet\_sourced}, (b) {\tt dangerous}.  Rows correspond to the error messages listed in Table \ref{tab:messages}.  Columns correspond to files.  In each matrix entry, white indicates no error, black indicates error.}
    \label{fig:matrix}
  \end{center}
\end{figure}

We constructed two relation matrices, one for {\tt internet\_sourced} (Figure \ref{fig:matrix}(a)) and one for {\tt dangerous} (Figure \ref{fig:matrix}(b)).  Each relation matrix has the same rows ($955$ distinct error messages, as shown in Table \ref{tab:messages}) but different numbers of columns ($9000$ for {\tt internet\_sourced} and $1000$ for {\tt dangerous}).

Continuing our example of the $1000$-th file in {\tt internet\_sourced}, this file corresponds to the $1000$-th column of the relation matrix in Figure \ref{fig:matrix}(a), and has $1$s in rows $58$, $254$, $393$, $589$, $683$, $910$ and $911$, because each of these messages occurred exactly once.  It has $0$s in all other entries in that column.

The horizontal dark bands present in both relation matrices shown in Figure \ref{fig:matrix} correspond to error messages that could not be categorized easily: not syntax errors or other specific malformations.  Some parsers produce these kind of messages more frequently than others, which explains why some portions of the matrices show a greater prevalence of dark horizontal bands than others.

Although there is some apparent structure visible in Figure \ref{fig:matrix}, namely the dark horizontal bands, it is difficult to discriminate any column-by-column differences.  These differences are indeed present, but require more sophistication to extract.  That statistical analysis forms the basis of most of this article.  

\section{Principal components analysis}
\label{sec:pca}

To build some intuition about the structure of the relation matrices, let us develop a dimension-reduced visualization of the columns (files) of both relation matrices shown in Figure \ref{fig:matrix}.  While there are many possible techniques for dimension reduction, principal components analysis is generally the easiest to construct and to interpret.  To better understand the structure of the data, we will incorporate the ground truth as part of the visualization.  This will help explain the performance of the Bernoulli misclassification test statistic in the next section.

Principal components analysis is a way to render a high dimensional data set that shows the ``most important'' dimensions and suppresses the rest.  It is therefore a convenient way to visualize data that are formatted as a set of points in $\mathbb{R}^n$ where $n$ is large.  The output of principal components analysis is a scatter plot in which the axes are chosen as the linear combinations of rows yielding the largest variance.  These axes are called the principal vectors, and are sorted from largest variance to least variance.  In our analysis, the largest three principal vectors were used because they represented the majority of the variance in the data.

To apply principal components analysis, we reinterpret our tabular data as a discrete subset of $\mathbb{R}^n$ (a point cloud) in which columns (files) are points, rows (messages) are components of the coordinates for each point. In both datasets, there are $n=955$ messages.  Because a file exhibits an error or not, the components are all either $0$ or $1$.  Although one might argue that this could result in quantization error, many interesting features are nevertheless visible in the two datasets.

\begin{figure}
  \begin{center}
   \includegraphics[width=5in]{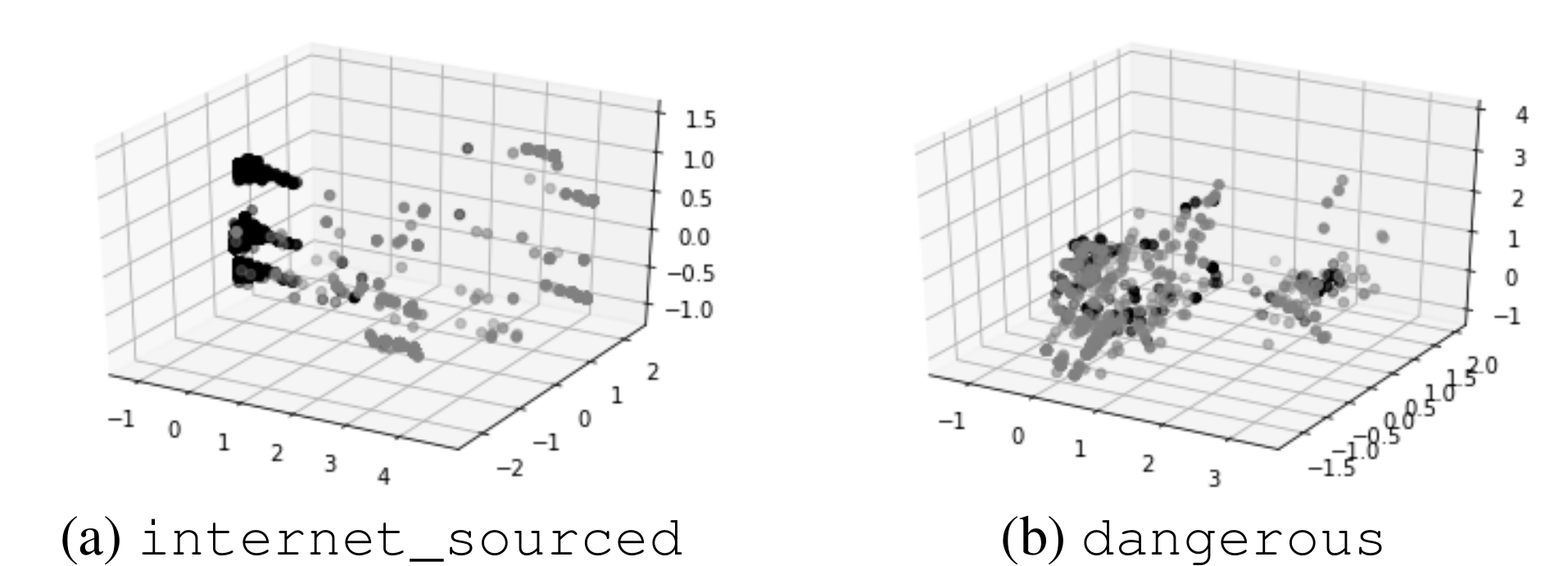}
    \caption{Principal components plots for (a) {\tt internet\_sourced} and (b) {\tt dangerous}.  Black indicate ``valid'', and gray indicates ``reject''.  The axes correspond to the three principal vectors, and so are not plotted on the same scale.}
    \label{fig:pca}
  \end{center}
\end{figure}

Figure \ref{fig:pca} shows the principal components analysis plots for both datasets.  Points in both plots are colored according to the ground truth so that a point corresponding to a ``valid'' file is black, and a point corresponding to a ``rejected'' file is gray.  The most striking difference in the principal components analysis plots is that the {\tt internet\_sourced} dataset is apparently much more ``clumpy'' than the {\tt dangerous} dataset.  The three dense clusters in Figure \ref{fig:pca}(a) consist entirely of ``valid'' files, with most of the ``rejected'' files in the {\tt internet\_sourced} data appearing as a ``haze'' of files outside of those clusters.  Although the cause of these three dense clusters of ``valid'' files cannot be determined solely from the relation matrix -- which files are accepted by which parsers -- we hypothesize that these clusters correspond to popular tool chains for creating PDF files.

In stark contrast, the {\tt dangerous} set shown in Figure \ref{fig:pca}(b) contains two loose clusters that are mixed ``valid'' and ``rejected'' files.  Intuitively, if one is looking to identify ``valid'' files, one would have a much harder time doing this with the {\tt dangerous} set, so we might argue that the apparent signal-to-noise ratio is much lower in the {\tt dangerous} set.

\begin{figure}
  \begin{center}
   \includegraphics[width=5in]{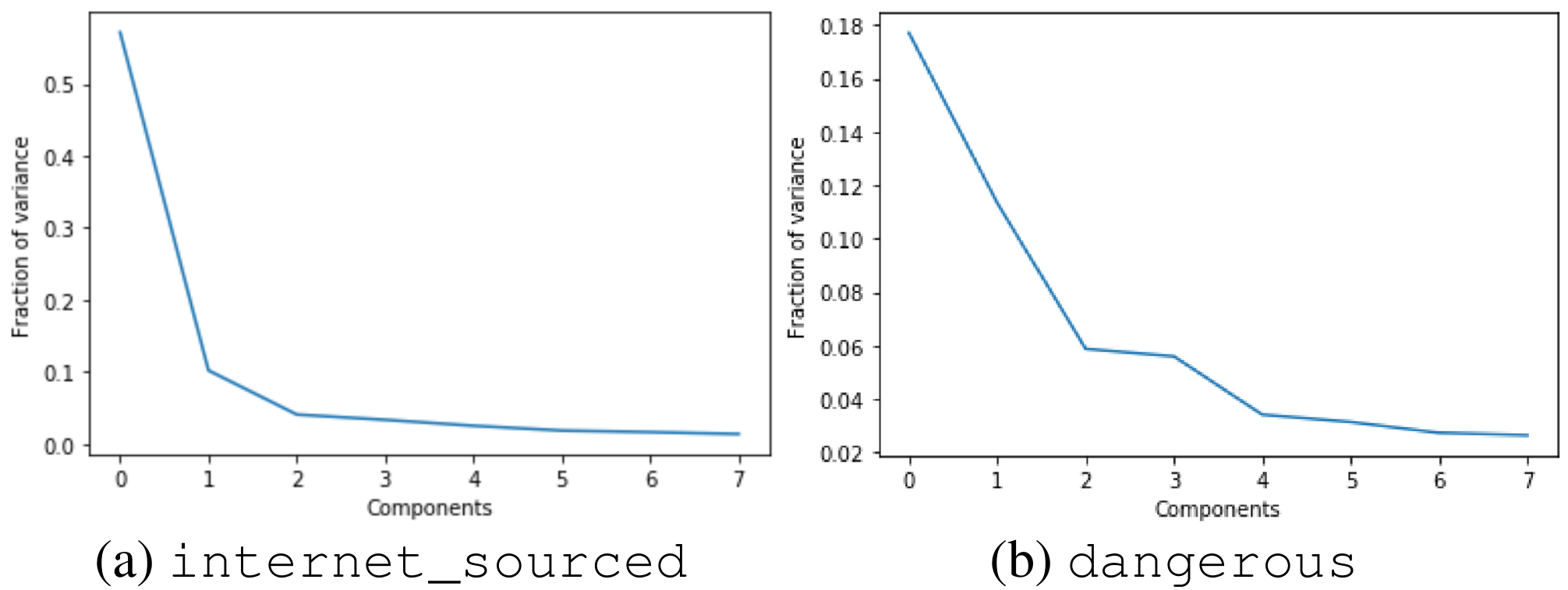}
    \caption{Scree plots for (a) {\tt internet\_sourced} and (b) {\tt dangerous}.}
    \label{fig:scree}
  \end{center}
\end{figure}

Principal components analysis can be misleading if only a small fraction of the overall variance in the data is explained by the first few principal vectors.  It is easy to determine if this problem is occurring -- simply plot the variance in the data explained by each principal vector.  This is called a Scree plot \cite{yong2013beginner}, and is shown in Figure \ref{fig:scree}.  In both datasets, the Scree plots decrease quite rapidly over the first few principal vectors. This shows that principal components analysis reliably represents the data.  

We can conclude that if one is generally working with files that are naturally occurring (like the {\tt internet\_sourced} set), one probably does not need to dig too deeply to determine whether a given file is valid or not.  The rest of this article buttresses this claim by providing a statistical test that does just that.  On the other hand, if one is routinely handling files that test the limits of their format (like the {\tt dangerous} set), statistical analysis alone will likely be insufficient to determine which files are valid.  A deeper format-aware analysis would be necessary in that case.

\section{Bernoulli likelihood ratio misclassification test statistic}
\label{sec:bern}

In order to determine file validity implicitly -- by consulting parser behavior rather than the specification itself -- we need exemplars of parser behavior on typical compliant and non-compliant files.  We propose to use the fact that the two datasets have rather different statistics (Table \ref{tab:datasets}) with the majority of files in {\tt internet\_sourced} being ``valid'' while a (slim) majority of files in {\tt dangerous} are ``rejected''.  We can attempt to identify files in {\tt internet\_sourced} that exemplify the parser behavior present in the {\tt dangerous} dataset -- these files probably should be treated with suspicion, and are perhaps not ``valid.''  Conversely, files within the {\tt dangerous} set that behave more like those in {\tt internet\_sourced} may well be ``valid.''  

A file is \emph{misclassified} if the set of messages it produces is more characteristic of the other dataset but not the one in which it is presently found.  We suspect that such a misclassified file will have a collection of error messages that is unusual when compared to the others in that dataset.  We estimate the probability that each error message occurs in each dataset, and estimate a likelihood for each file assuming the error messages are independent.  Since both datasets have the same error messages, we can \emph{also} estimate the likelihood that the file came from the other dataset.  A likelihood ratio statistic can thereby detect when a file is misclassified, because it is more likely to belong to the other dataset.

Since we cannot assume that the occurrence of any given set of error messages is statistically independent, it is difficult to write a proper likelihood function.  To that end, we use a \emph{pseudo-likelihood}, which makes the incorrect assumption that error messages are statistically independent \cite{arnold1991pseudolikelihood}.  On the other hand, this assumption merely reduces the sensitivity of the test without producing extra outliers.  The pseudo-likelihood assumption trades \emph{recall} to get better \emph{precision}.  As such, pseudo-likelihoods are useful for classification but not for uncertainty quantification.

For the Bernoulli misclassification test, we assume each error message is governed by a Bernoulli distribution, which means that it either occurs or does not occur.  Multiple instances of the same error are ignored.  The test assumes that each error message occurs with a probability that depends on the dataset (either {\tt internet\_sourced} or {\tt dangerous}).  When a parser processes a given file, sometimes it produces multiple copies of the same error message.  This can happen if the parser attempts to repair a slightly non-compliant file as it proceeds, for instance.  If this happens, then the Bernoulli distribution is no longer valid because it assumes at most one instance of a given error message.  Ultimately, the performance of the Bernoulli misclassification test was good even though we ignored duplicate error messages.

\begin{figure}
  \begin{center}
   \includegraphics[width=5in]{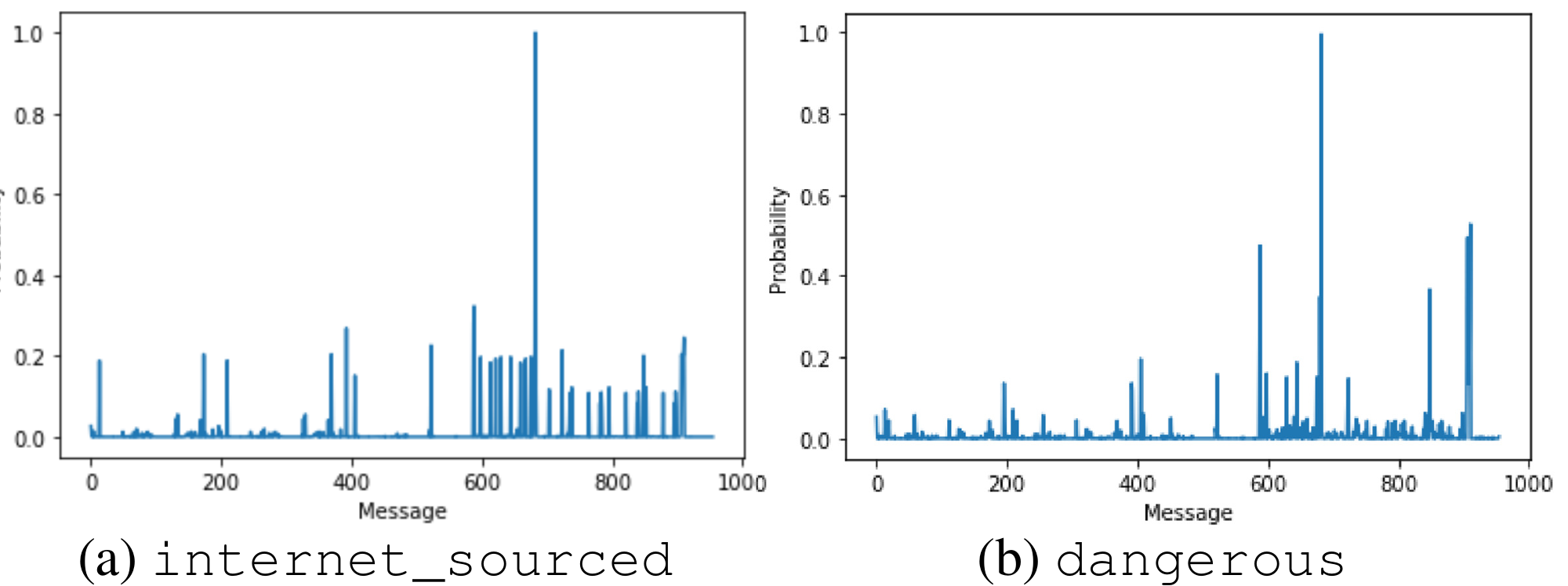} 
    \caption{Error probability for (a) {\tt internet\_source} and (b) {\tt dangerous}.}
    \label{fig:error_prob}
  \end{center}
\end{figure}

Let us consider the {\tt internet\_sourced} dataset first.  It is straightforward to compute the probability $p_k$ of error message $k$ occurring from the relation matrix: simply take the average value of row $k$ in the relation matrix shown in Figure \ref{fig:matrix}(a).  The resulting probabilities for both datasets are shown in Figure \ref{fig:error_prob}.  Said another way, if file $f$ is drawn from the {\tt internet\_sourced} dataset, then the probability that $f$ produces error message $k$ is $p_k$.  Conversely, the probability that $f$ does \emph{not} produce this error message is $(1-p_k)$.  If we let $f_k = 0$ if the file $f$ did not produce error $k$ and $f_k=1$ if the file did produce error $k$, then the probability that $f$ is from the {\tt internet\_sourced} dataset is
\begin{equation*}
  p_k f_k + (1-p_k)(1-f_k),
\end{equation*}
if we only consider error message $k$.

\begin{figure}
  \begin{center}
   \includegraphics[width=5in]{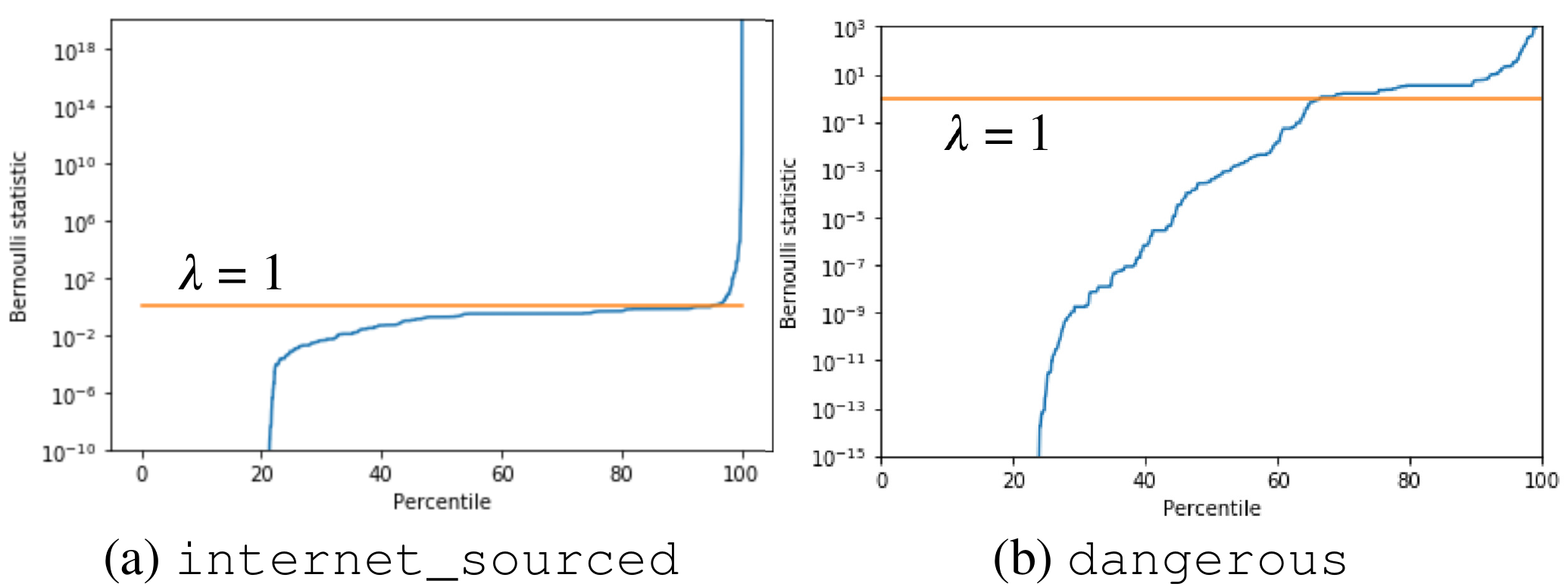}
    \caption{Histogram of Bernoulli misclassification test statistic for (a) {\tt internet\_sourced} and (b) {\tt dangerous}.  The horizontal line marks a value of $\lambda=1$.}
    \label{fig:ll_bern}
  \end{center}
\end{figure}

Since we have many error messages available for analysis, the pseudo-likelihood that file $f$ (column in the relation matrix) is correctly classified is simply the product of each of these individual probabilities, namely
\begin{equation*}
  L_{\tt internet\_sourced}(f) = \prod_{k=1}^{955} (p_k f_k + (1-p_k)(1-f_k)).
\end{equation*}
We define $L_{\tt dangerous}(f)$ similarly using the error message probabilities $p'_k$ from the {\tt dangerous} set instead,
\begin{equation*}
  L_{\tt dangerous}(f) = \prod_{k=1}^{955} (p'_k f_k + (1-p'_k)(1-f_k)).
\end{equation*}
We define the \emph{Bernoulli misclassification test statistic} to be the ratio of these two pseudo-likelihoods,
\begin{equation*}
  \lambda_{\tt internet\_sourced}(f) = \frac{L_{\tt dangerous}(f)}{L_{\tt internet\_sourced}(f)}.
\end{equation*}
If $f$ is a file drawn from the {\tt internet\_sourced} dataset, then we generally expect that $L_{\tt internet\_sourced}(f)$ will be larger than $L_{\tt dangerous}(f)$, which implies that $\lambda_{\tt internet\_sourced}(f)<1$.
Conversely, if a file is drawn from the {\tt dangerous} dataset, which means it is a misclassification if it actually is present in {\tt internet\_sourced}, we would expect that $\lambda_{\tt internet\_sourced}(f)>1$.
The intuition is that since files in the {\tt dangerous} set are likely to be invalid, a high value of $\lambda_{\tt internet\_sourced}(f)$ is an indication that the file $f$ is not compliant.
A histogram of $\lambda_{\tt internet\_sourced}$ is shown in Figure \ref{fig:ll_bern}(a).

Conversely, we can define
\begin{equation*}
  \lambda_{\tt dangerous}(g) = \frac{L_{\tt internet\_sourced}(g)}{L_{\tt dangerous}(g)}
\end{equation*}
for each file $g$ in the {\tt dangerous} set.  The intuition in this case is that a high value of $\lambda_{\tt dangerous}(g)$ is an indication that $g$ is compliant, since it is likely to be a misclassification.  The histogram of values of $\lambda_{\tt dangerous}(g)$ is shown in Figure \ref{fig:ll_bern}(b).

Since there is some variability (or noise) present within the data, we should not use the value of $\lambda = 1$ as the cutoff for detecting misclassifications.
Although the intersections between the histogram curves and the red lines $\lambda=1$ in both frames of Figure \ref{fig:ll_bern} are close to the true fraction of misclassifications in both datasets ($80\%$ for {\tt internet\_sourced} and $48\%$ for {\tt dangerous}), they are not exactly correct.
This suggests using a different threshold $T$ instead, so that all files whose statistic $\lambda$ is greater than $T$ will be deemed to be misclassified.

Let us now use the ground truth, which specifies whether a given file is ``valid'' or  ``rejected'', to determine the performance of the Bernoulli misclassification test statistic.
A misclassified file in {\tt internet\_sourced} is one that is ``rejected'', while a misclassified file in {\tt dangerous} is one that is ``valid''.

In the case of either dataset, for a given threshold $T$, the \emph{probability of detection} is the probability that a truly misclassified file $f$ will have a statistic $\lambda(f)>T$.
On the other hand, the \emph{probability of false alarm} is the probability that a correctly classified file $f$ will have a statistic $\lambda(f)>T$.
Ideally, the probability of detection will be close to $1$, while simultaneously the probability of false alarm will be close to $0$.

\begin{figure}
  \begin{center}
   \includegraphics[width=5in]{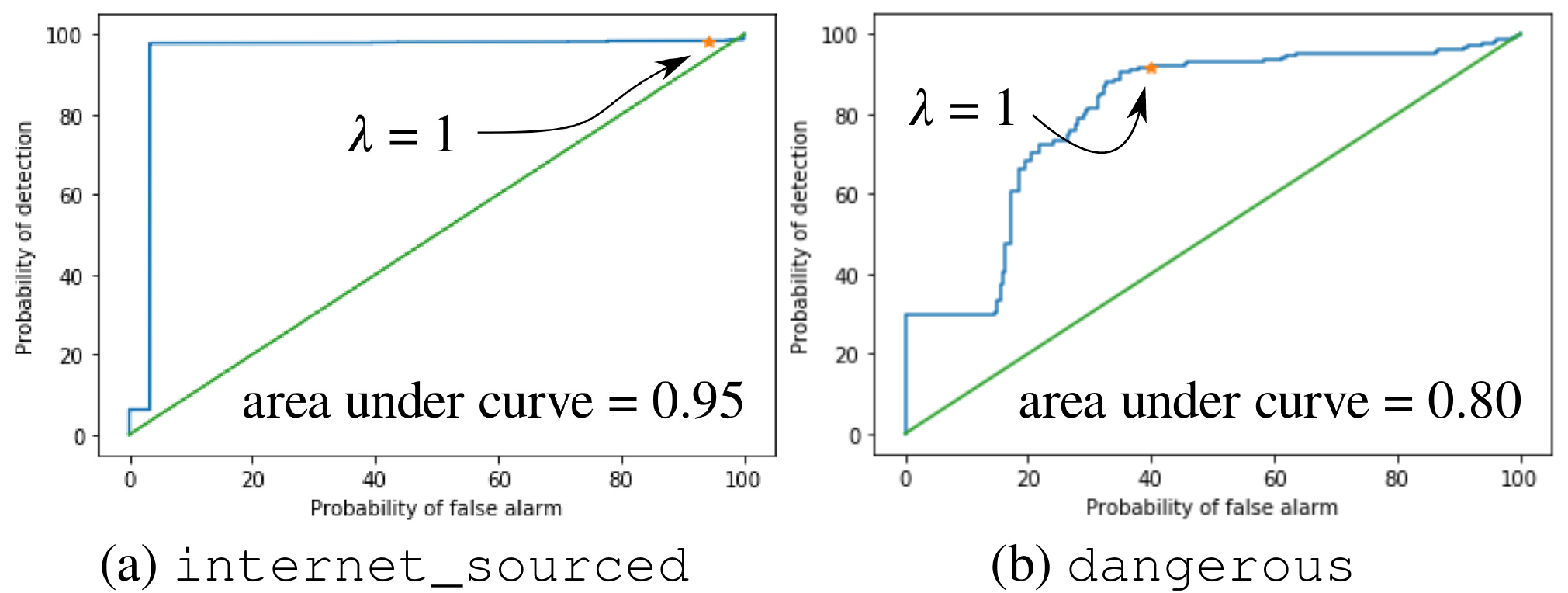}
    \caption{Receiver operating curves for Bernoulli likelihood ratio misclassification statistic for (a) {\tt internet\_sourced} and (b) {\tt dangerous}.}
    \label{fig:roc_bern}
  \end{center}
\end{figure}

The plot of probability of detection versus false alarm for all thresholds is shown in Figure \ref{fig:roc_bern}.
Better misclassification detectors have plots further to the upper left, away from the diagonal.
Since the curve plotted is far above the diagonal for {\tt internet\_sourced} in Figure \ref{fig:roc_bern}(a), we conclude that the Bernoulli likelihood ratio misclassification statistic is a very accurate detector of misclassified files in {\tt internet\_sourced}.
Additionally, since the plot in Figure \ref{fig:roc_bern}(b) is above the diagonal, this indicates that the Bernoulli likelihood ratio misclassification statistic also performs well on the {\tt dangerous} dataset set.  The sharp plateau in Figure \ref{fig:roc_bern}(b) is due to a number of instances in which $L_{\tt internet\_sourced}$ took the value $0$ on some files in {\tt dangerous}.  These happen to be truly non-compliant files!
These qualitative assessments are confirmed by the areas under the curves in Figure \ref{fig:roc_bern}.  A perfect misclassification detector will have an area of $1$ under the curve, while a detector that randomly decides misclassifications would have area $0.5$ under the curve.  We obtained $0.95$ for $\lambda_{\tt internet\_sourced}$ and $0.80$ for $\lambda_{\tt dangerous}$ for areas under the curve, which we deem to be surprisingly good given the fact that the file contents were not directly inspected by our method.

\section{Parser redundancy analysis}

The Bernoulli misclassification statistic relies upon the diversity of responses to each file in order to make reliable decisions.
It is impractical to use quite so many parsers as we used, so it is useful to assess whether we could get good performance with fewer parsers.
The easiest way to do this is to measure the correlation between the typical error message distributions produced by different parsers across all files.
In the following analysis, we combined both {\tt internet\_sourced} and {\tt dangerous} into a single dataset,
whose relation matrix consists of the horizontal concatenation of both matrices shown in Figure \ref{fig:matrix}.
That is, the rows are the same as before, but the columns consist of the union of the columns from both matrices.
Error messages (rows) that never occur were removed from the analysis, since they contribute no information.   

\begin{figure}
  \begin{center}
   \includegraphics[width=5in]{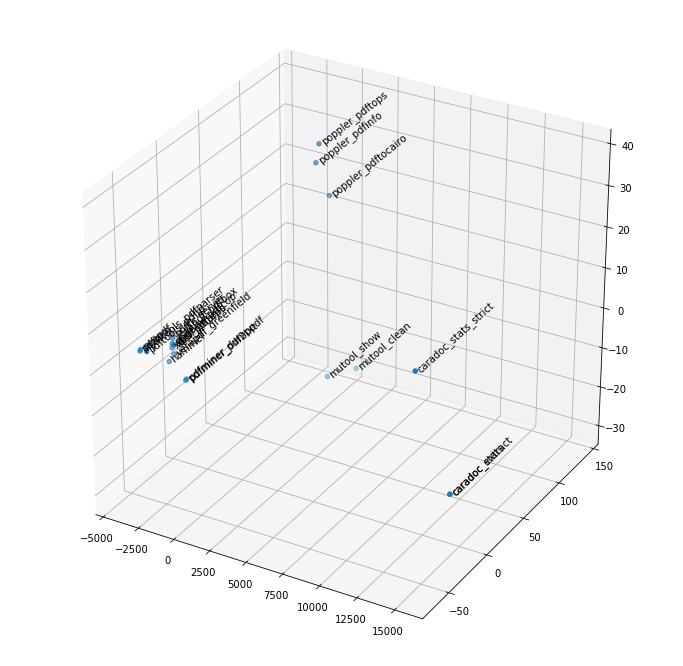}
    \caption{Parsers placed in three dimensions according to principal components analysis of the union of {\tt internet\_sourced} and {\tt dangerous}.}
    \label{fig:parser_pca}
  \end{center}
\end{figure}

The data can also be visualized using principal components analysis, much as was done in Section \ref{sec:pca}, but instead we use the error counts across files as coordinates for each parser.  Since specific error messages cannot be enabled or disabled individually, it makes sense to aggregate the error messages into parsers by taking the total number of error messages for a each parser on a given file.  Principal components analysis places the parsers according to the plot shown in Figure \ref{fig:parser_pca}.
There is one large cluster (in the lower left of Figure \ref{fig:parser_pca}) of parsers which have similar behaviors.
There are quite a few outliers, most notably {\tt caradoc\_extract} and {\tt poppler\_pdfinfo}.  The reader is cautioned that the apparent density of the large cluster is a bit misleading, because the ranges of the axes are quite different.  In any event, the wide spread of parsers across the plot indicates that having a large number of parsers is quite valuable.

\begin{figure}
  \begin{center}
   \includegraphics[width=5in]{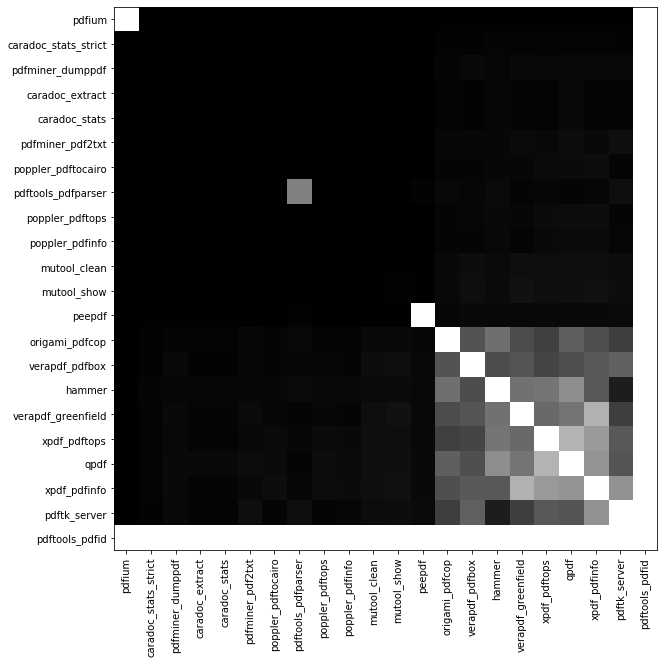}
    \caption{Correlation matrix between parser behaviors.  Rows and columns are sorted according to median message correlation.}
    \label{fig:parser_correlation}
  \end{center}
\end{figure}

Although principal components analysis is useful, we can obtain a ranking of parsers by their redundancy, so that we can prioritize more informative parsers.
We computed Pearson's correlation coefficient between all pairs of error messages (rows), to form a fairly large correlation matrix (not shown).
Different error messages that occur on exactly the same set of files have a correlation coefficient of $1$.  In that case, one of those error messages is redundant.
Errors were then grouped by parser, and for each pair of parsers we stored the median correlation from each of their pairwise message correlations.
The result of this aggregation is the matrix shown in Figure \ref{fig:parser_correlation}.  Whiter colors indicate higher correlation -- more redundancy -- while darker colors indicate lower correlation.  Most of the matrix is fairly dark, which indicates low redundancy overall.  The bright off-diagonal entries indicate trade-offs: one needs only run one of the two parsers indicated.   The bright bands for {\tt pdftools\_pdfid} occurred because that parser did not correlate with any of the others at all.

The clusters visible in the Figure \ref{fig:parser_pca} can be confirmed by sorting the parsers based on their median correlation.  The rows and columns shown in Figure \ref{fig:parser_correlation} are sorted in this way.  This indicates which parsers are individually most informative, because they form an approximate spanning set for the data.   The obvious block in the lower right suggests that parsers should be grouped into two categories based on their informativeness: high and low.  The rows and columns of the block in Figure \ref{fig:parser_correlation} suggest that the boundary between the high and low categories appears to be fairly sharp, with all parsers to the left of {\tt origami\_pdfcop} being highly informative.  

Based on the approximate spanning property of the highly informative category, which covers most of the variability visible in Figure \ref{fig:parser_pca}, one should always run the parsers in the high category, with the other parsers in the low informative category parser treated as increasingly optional as one moves to the lower right of Figure \ref{fig:parser_correlation}.  Notice that the parsers near the lower right of Figure \ref{fig:parser_correlation} all come from the cluster in the lower left of Figure \ref{fig:parser_pca}.  Although it is difficult to see from the projection shown in Figure \ref{fig:parser_pca}, least informative parsers are on the \emph{inside} of the cluster, while the parsers in the same cluster that are on the outer edges of the cluster are more informative.

\section{Conclusion and recommendations}

This article has demonstrated that a given file's compliance with a format specification can be determined from the following three samples: (1) a diverse collection of parsers for the file format, (2) a sample of compliant files, and (3) a sample of non-compliant files.  Our methodology is format agnostic, and so could work if a file format is not formally specified.  Furthermore, our Bernoulli test for compliance is statistical, so it is both robust to errors in the three samples, yet benefits from richer samples should they be available.  We note that the use of the three samples means that this approach is a supervised approach.  Under appropriate circumstances, it might be possible to use an unsupervised bootstrapping approach to extract the two samples of files from a single aggregated sample.  This requires further investigation.

Principal components analysis is helpful in understanding the variability within a sample of files or parsers, but it holds somewhat less value as an automated analytic tool.
Clusters visible within the principal components plots appear to reflect specific tool chains used in the creation of files, with valid files forming several dense clusters.

The Bernoulli likelihood ratio statistic detects non-compliant files by comparing error message prevalence aggregated across all parsers and both samples of files.  It relies inherently upon both the coverage and depth of these samples, but we have shown that it can be very effective at its job when these samples are adequate.

Overall, we found that there is not much redundancy present in the behaviors of parsers for the PDF specification.  That is to say that the relation matrix contains by far the most information if all parsers are considered, so one ought to use all parsers if possible.  Our analysis determined which parsers are individually the most informative, based on pairwise comparisons.  While it is entirely reasonable to study different subsets of parsers, we did not perform that analysis here.  We refer the interested reader to \cite{ambrose2020topological} where such an analysis was performed.

If resources are tight, we found that it is probably not necessary to rerun a given parser with different options.  For instance, running only one of {\tt xpdf\_tops} and {\tt xpdf\_pdfinfo} probably will not change the results too much.  Roughly speaking, it appears that the best strategy is ``more programmers contributing different code instead of one programmer's code run in many different ways.''

Our two statistical techniques for analyzing file format compliance are admittedly simple but standard statistical tools, though are apparently not in wide use.  They are easy to deploy, easy to interpret, and require little maintenance other than the selection of a single detection threshold.  We therefore encourage deeper exploration into and experimentation with statistical methods in the study of file format compliance.

\section*{Acknowledgments}
The author would like to thank the SafeDocs test and evaluation team, including NASA (National Aeronautics and Space Administration) Jet Propulsion Laboratory, California Institute of Technology and the PDF Association, Inc., for providing the test data.  The author would like to thank Kris Ambrose for his heroic processing of all of the relevant files through all of the parsers, and for subsequently packaging the results in a very convenient format.  The author would also like to thank Peter Wyatt for his detailed and insightful investigation into the files that were deemed outliers and for closely reading an earlier draft of this article.  

This material is based upon work supported by the Defense Advanced Research Projects Agency (DARPA) SafeDocs program under contract HR001119C0072.  Any opinions, findings and conclusions or recommendations expressed in this material are those of the author and do not necessarily reflect the views of DARPA.

\bibliographystyle{plainnat}
\bibliography{docstat_bib}
\end{document}